# Design principles for textured multi-layered composites using magnetically assisted slip casting

Hortense Le Ferrand, Florian Bouville*, André Studart*

*Complex Materials, Department of Materials, ETH Zurich, 8093 Zurich, Switzerland*

*florian.bouville@mat.ethz.ch, *andre.studart@mat.ethz.ch

**Abstract**

In many natural multi-layered composites, such as in the dactyl club of Stomatopods or the shell of bivalve mollusks, the defining functional and structural properties are determined by locally varying orientations of inorganic building blocks within each layer. One approach to artificially produce textured microstructures inspired by such complex composites is magnetically assisted slip-casting (MASC). MASC is a colloidal process in which anisotropic particles are oriented at arbitrarily defined angles using a magnetic field. The orientation of the particles is maintained during growth as particles are collected from the wall of a porous mold. Whereas a number of proof-of-concept studies have established the potential of the technique, the full design space available for MASC-fabricated structures, and the limits of the approach, have so far not been explored in a systematic manner. To fill this gap, we have studied both theoretically and experimentally the various torques that act on the particles at different stage of the assembly process. We define the boundary conditions of the MASC process for magnetically responsive alumina platelets suspended in a low-viscosity aqueous suspension, considering the composition of the colloidal suspension and the dynamics of the particle-alignment process under a rotating magnetic field. Taken together, these findings define a design strategy for the fabrication of designed multi-lamellar microstructures using MASC. Our guidelines are based on a comprehensive understanding of the physical mechanisms governing the orientation and assembly of anisotropic particles during MASC and indicate a route to expanding this technique to building blocks of various chemistries, and thus to broadening the range of bio-inspired composites with customized multi-scale structures that can be produced for specific applications.

**Introduction**

Natural composites possess unique complex structural features spanning across several length scales[1]. Evolution guided organisms towards producing hierarchical structures with outstanding properties[2,3]. In particular, natural periodic assemblies with decreasing pitch dimensions, such as Bouligand or helicoidal structures found for instance in beetles and stomatopods, display broadband phononic and/or photonic properties[4,5]. Reproducing natural periodic textures in organic–mineral composites with a similar degree of control is highly attractive, with a view to creating materials with novel functionalities. For example, reproducing the helicoidal microstructure of stomatopods could enable the unusual combination of properties such as hardness and toughness while other functionalities can be provided by the intrinsic nature of the building blocks, for instance thermal conductivity. Growing such composites, however, poses substantial challenges due to the hindered mobility of particles when suspended in dense suspensions[6].

The building blocks of natural multi-layer composites are typically stiff anisotropic particles of micrometer dimensions. These units are assembled in controlled processes guided by the structure and function of the final construction. The spatial texture of the composites can take the form of brick-and-mortar, prismatic, cross-lamellar and plywood structures[1]. Various bottom-up colloidal assembly techniques have been utilized to create multi-layered textured composites inspired by natural materials. Self-assembly at water–air interfaces, layer-by-layer spin coating, sedimentation in dilute conditions or tape casting have been shown to be efficient methods for aligning the platelets almost perfectly in the horizontal plane[7–10], whereas vertical alignment can be generated using freeze-casting[11]. To orient anisotropic particles into arbitrary direction, the use of electrical and magnetic fields has been explored[12,13].

All of the above-mentioned methods, however, are limited to a single orientation within a bulk specimen. To address this limitation, we have recently introduced the magnetically assisted slip-casting (MASC) technique[14], a one-step process that enables the controlled continuous variation of platelet orientation at different locations in centimeter-scale composites. MASC is based on the well-established slip-casting process, where an aqueous suspension of spherical colloidal particles is casted onto a porous substrate, typically made of gypsum. During slip casting, removal of water from the suspension through the pores of the gypsum concentrates and therefore consolidates the particle deposit into a jammed layer. In MASC, the spherical particles are replaced with micrometric platelets characterized by ultra-high magnetically response (UHMR)[13]. The orientation of those micrometric platelets can be controlled during the growth of the deposit using programmed rotating magnetic fields that change directions over time, leading to the formation of unique periodic or semi-periodic biomimetic microstructures.

The potential of the MASC approach has been explored in a number of application areas[14,15], but a comprehensive exploration of the design space for fabricating complex structures with desired textures requires a fuller understanding of the physical principles underlying the technique. Specifically, MASC assembly depends on the temporal response of the UHMR anisotropic particles, the kinetics of water removal during conventional slip casting and on the various forces acting on the particles during casting. Whereas the temporal response of UHMR anisotropic particles and the kinetics of water removal through porous substrates have been described separately in the literature[16–18], the forces acting during combined magnetic alignment and slip casting remain to be studied in a methodical manner.

Here we establish the boundary conditions and the processing window of the MASC technique by determining the forces and torques applied on microplatelets at each stage of the assembly process. Based on theoretical predictions and experimental data, we first explored how to tailor the composition of the initial suspension of anisotropic particles to obtain a growing jammed layer with an arbitrarily defined angle of alignment after assembly and drying. Once the ideal composition of the initial suspension was determined, we studied the dynamics of the particle-alignment process in this concentrated system. To determine the minimum time needed to rotate platelets in a given alignment plane for different rotation directions and magnetic-field strengths, the particle motion in response to a variation in magnetic-field direction was measured using *in situ* optical imaging. This enabled us to determine the minimum times needed to rotate platelets in a given alignment plane for different rotation directions and magnetic-field strengths. The quality of the alignment in monolayers and multiple MASC-fabricated layers was quantified using x-ray diffraction. Finally, we produced several example structures to illustrate the sort of periodic composites that can be produced based on our broadened understanding of the MASC method.

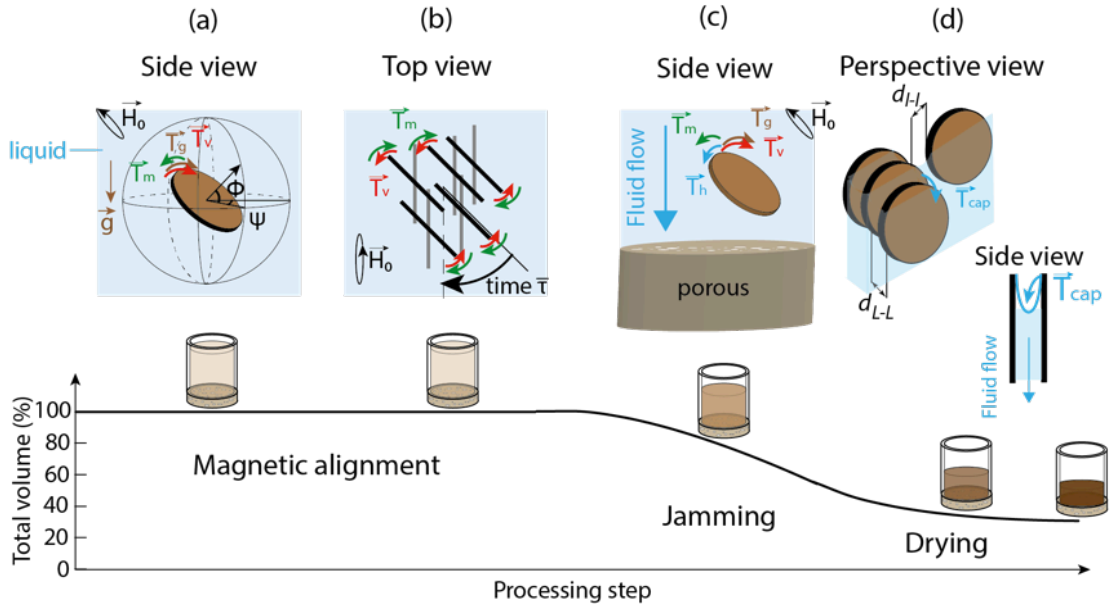

**Figure 1: Processing steps and torques involved during the MASC process. (a)** Initially, anisotropic platelets are suspended in a low viscosity liquid and submitted to the gravitational field $\vec{g}$, a rotating magnetic field $\overrightarrow{H_0}$ and the viscous drag imposed by the liquid. The balance between those three torques, named $\overrightarrow{T_g}$, $\overrightarrow{T_m}$ and $\overrightarrow{T_v}$ respectively, defines the orientation of the platelet in space, described by the two angles $\phi$ and $\psi$. **(b)** To vary the orientation of the platelets between each layer, the rotating magnetic field is varied dynamically. When the aligned platelets are submitted to a new rotating magnetic field, they need a time $\bar{\tau}$ to align along the new plane of orientation. In the case described in the cartoon, only $\overrightarrow{T_m}$ and $\overrightarrow{T_v}$ are acting because we consider only a variation of angle $\psi$, the platelets remaining vertical. **(c)**

During MASC, the liquid is slowly removed through the pores of the substrate. This induces a fluid flow that adds an additional hydrodynamic torque $\vec{T_h}$ to the platelet. Removal of the fluid leads to the jamming of the suspension. **(d)** The final step of the process involves complete drying of the jammed cake. The flow of the liquid between the aligned platelets generates capillary torques $\vec{T_{cap}}$ which amplitudes depend on the distance between the aligned platelets.

**Theoretical description of the torques during the MASC process**

At each stage of the MASC process, different torques of varying intensities influence the alignment of platelets and eventually control the microstructure in the final assembled dried structure or green body. MASC is based on the suspension of magnetically responsive anisotropic particles in a low-viscous medium and the subsequent control of the particles orientation by a rotating magnetic field. As in conventional slip-casting processes, the suspension is poured on top of a porous mold, typically gypsum. Capillary forces associated with the porous structure drive the slow removal of the solvent, leading to a steadily increasing particle concentration. The removal of the solvent introduces hydrodynamic torques on the particles that needs to be considered as they can rotate and orient the anisotropic particles. During the casting step, the particle orientation can be controllably changed by applying a magnetic torque, a degree of freedom that is specific to the MASC process. The ultra-fast magnetic temporal response of micrometric platelets submitted to a rotating magnetic field, ranging between 100 ms to a few s, has already been described in diluted suspensions (up to 5 vol% of platelets)[17,18], and without time-dependent removal of the fluid. The MASC method, however, typically involves both up to eight times higher volume fractions of particles (40 vol%) and fluid removal. Finally, at the end of the MASC process, a capillary torque is introduced during the final drying step by the air–water interface within the sample. This drying step is crucial as the resulting assembly needs to resist the capillary forces arising from the moving air–water interface that will tend to make the assembly collapse.

As the different torques vary in intensity during the process (see Fig. 1), each stage — that is, magnetic alignment in suspension, fluid removal and jamming, and final drying — should be treated separately. The individual torques expression can be derived from previous studies conducted in dilute suspension: the gravitational torque $\vec{T_g}$, the magnetic torque $\vec{T_m}$ from the rotating magnetic field $\vec{H_0}$ and the viscous torque $\vec{T_v}$ from the surrounding fluid from Erb *et al.*[18], whereas the hydrodynamic torque $\vec{T_h}$ arising from the flow of fluid during its removal from [16,19] and $\vec{T_{cap}}$ from [20,21]. However, the interplay between the forces in concentrated suspensions and their effect on the dynamics of the MASC process has not yet been studied and is required to control accurately the final microstructure.

*Stage 1: magnetic alignment in suspension*

In the concentrated suspension, the platelets experienced $T_g$, $T_m$, and $T_v$, which can be expressed as from dilute suspensions if there is not steric interaction between them: the suspension is liquid enough.

To simplify the writing of those torques, we consider their expression in the configuration where they are at their maximum:

$$T_g = \pi \frac{lL^3}{8} \rho g \tag{1}$$

$$T_m = \frac{2\pi \mu_0 \chi_{ps}^2}{3(\chi_{ps}+1)} \left[ \left(\frac{l}{2}+k\right)\left(\frac{L}{2}+k\right)^2 - \frac{lL^2}{8} \right] \cdot H_0^2 \tag{2}$$

$$T_v = \frac{3}{2} \pi^2 lL^2 f \eta \tag{3}$$

where $\rho$ is the density of the particles, $l$ and $L$ the thickness and diameter of the particles respectively, $\mu_0 = 4\pi \cdot 10^{-7}$ N.A$^{-2}$ the permeability of free space, $\chi_{ps}$ the volume susceptibility of the particles, $f = \frac{1+\frac{0.64}{\ln(2s)}}{1-\frac{0.5}{\ln(2s)}}$ is the geometrical friction factor for a platelet of aspect ratio $s = \frac{L}{l}$ and $\eta$ the viscosity of the Newtonian surrounding fluid.

From these expressions it is clear that increasing magnetic susceptibility or decreasing viscosity will facilitate platelet alignment. However, the time needed for complete platelet alignment cannot be determined directly from these formulae and should still be assessed.

*Stage 2: Liquid removal and jamming*

During slip casting, capillary forces and the resulting pressure gradients drive the liquid to flow from the suspension into the mold. This volumetric flow, *Q*, generates drag forces on the particles, which induce orbital motion that can be described by Jeffery orbits[22,23]. Once the liquid is removed from the suspension, a concentrated deposit of particles forms at the surface of the mold. As casting proceeds, the fluid from the suspension has to go through this newly formed jammed particle layer. As this layer has some tortuosity, the resistance to the fluid flow increases with the growth of the deposit, resulting in a pressure decrease[24]. As a consequence, the volumetric flow *Q* decreases with time *t* and distance *x* from the surface *S* of the substrate [16,19]:

$$|Q(t)| = |\dot{V}| = S \left|\frac{dx}{dt}\right| \tag{4}$$

For the present example, we consider a cylindrical geometry of a vertical plastic tube glued on top of a gypsum substrate. In that case, the volumetric fluid flow during slip casting is given by the expression

$$|Q(t)| = \pi R^2 |\dot{x}| = \frac{\pi R^2}{2} \frac{a}{\sqrt{t}} = \frac{\pi R^2}{2} \frac{a^2}{x} \tag{5}$$

where *R* is the diameter of the cylindrical casting tube and $x(t) = a\sqrt{t}$ describes the growth of the deposit. The constant $a$ depends on system conditions such as the particle concentration, which controls the tortuosity of the jammed layer, and the hydrostatic resistance $\delta$, the capillary pressure $\Delta p$ at the pores of the mold and $J$ the ratio between the volume of the deposit and that of the extracted liquid,

$$a = \sqrt{\frac{2J\Delta p}{\eta\delta}} \tag{6}$$

The maximum hydrodynamic torque experienced by the platelets, assuming no inter-particles interactions, is[22]

$$T_h = \frac{8\pi L^3 f}{3\ln(2s)}\eta \frac{16}{\pi R^4} Q \tag{7}$$

By combining equations (7) and (5), we obtain the maximum hydrodynamic torque experienced by the platelets as a function of the thickness $x$ of the deposit:

$$T_h = \frac{8\pi L^3 f}{3\ln(2s)}\eta \frac{16}{\pi} \frac{\pi R^2}{4} \frac{a^2}{2} \frac{a^2}{x} = \frac{8\pi L^3 f}{3\ln(2s)}\eta \frac{8}{R^2} \frac{a^2}{x} \tag{8}$$

From these theoretical expressions, it can be seen that the hydrodynamic drag torque dominates the platelets orientation in the deposit formed close to the mold surface, but is negligible above a critical deposit thickness. $Q$ is also dependent on the tortuosity of the cake layer and can be changed therefore by varying the amount of particles and their packing.

*Stage 3: Drying*

When a meniscus forms between two platelets separated by a distance $d$, the Laplace pressure $\Delta P$ increases leading to capillary forces directed towards the mold's surface. The resulting torque $T_{cap}$ that drives a platelet from a vertical to a horizontal orientation is then

$$T_{cap} = \Delta P \cdot L^2 \cdot \left(L + \frac{d}{2}\right) = \frac{2\gamma\cos\beta}{d} \cdot L^2 \cdot \left(L + \frac{d}{2}\right) = \frac{2\gamma L^2}{d} \tag{9}$$

where $\gamma$ is the interfacial energy between the platelet and the solvent, $d$ the inter-platelet distance as depicted in Fig. 1C, and $\beta$ the angle of the meniscus formed with the platelet [21,22]. Experimental measurements of $d$ as a function of the volume fraction are shown in a Supplementary Figure and give a value of $d$ comprised between 12 µm and 100 nm. For any positive value of $\gamma$ and d below 12 µm the inequality $T_{cap} > T_g$ is always fulfilled. This means that the torque applied by the moving air–water interface is always larger than the torque necessary to rotate the platelets from vertical to horizontal position. The platelets can only remain in their vertical position, despite the capillary torque, when an assembly is formed that is sufficiently strong to block the movement of individual platelets, similar to a house-of-cards structure. To ensure that the neighbouring assembly disables platelet movements, the free volume in which the particle can rotate needs to be smaller than the hydrodynamics radius of the platelets, which is by definition the equivalent radius within which an individual platelet is free to rotate. The following condition needs therefore to be fulfilled:

$$V_{free} < V_{hydrodynamics} \tag{10}$$

where $V_{hydrodynamics} = \frac{4}{3}\pi R_h^3$, $\tag{11}$

$R_h$ the hydrodynamic radius calculated using the following formula[25]:

$$R_h = \frac{3L}{4} \frac{1}{\sqrt{1+\left(\frac{l}{L}\right)^2} + \frac{L}{l}\ln\left(\frac{l}{L} + \sqrt{1+\left(\frac{l}{L}\right)^2}\right) - \frac{l}{L}}. \quad (12)$$

and $V_{free}$ is free volume expressed as the volume of fluid around one particles and can be written as a function of the volume fraction of particle in suspension as:

$$\nu = \frac{V_p}{V_{free}} \quad (13)$$

with $V_p = \frac{\pi}{4} \cdot L^2 \cdot l$ the volume of a disc-shape platelet.

Thus $V_{free}$ is equal to:

$$V_{free} = \frac{\pi \cdot L^2 \cdot l}{4\,\nu} \quad (14)$$

Therefore, the minimum volume fraction $\nu_{min}^{opt}$ for which the particles maintain their orientation is obtained at $V_{free} = V_{hydrodynamics}$ and $\nu_{min}^{opt} = \frac{3}{16} \frac{L^2 \cdot l}{R_h^3}. \quad (15)$

On the other end of the volume-fraction range, the maximum concentration at which the particles can still rotate and align with the magnetic field corresponds to the jamming concentration where platelets form a percolating network.

## Materials and Methods

**Magnetized platelets**

Alumina platelets of aspect ratios 30 (Ronaflair, Whitesapphire) were provided by Merck, Germany; they are 300 ± 130 nm in thickness and 8.9 ± 2.9 µm in diameter. They were magnetized using superparamagnetic iron oxide nanoparticles (SPIONs, EMG-705 Ferrotec, Germany) following the procedure reported in ref. 13, where the same 0.1-% volume fraction of SPIONs to alumina platelets was used.

**Suspension preparation and rheological characterization**

The platelets were dispersed in water to reach the desired volume fraction, with the help of a surfactant, Dolapix (CE 64, Zschimmer & Schwarz, Germany) and ultrasonication for 5 min at 50 % of the total power (UP200s, Hielscher, Germany). 5 wt% of polyvinylpyrroliydone (PVP, MW = 360 000, Sigma-Aldrich, Switzerland) was added and mixed with a magnetic stirrer for an hour before further ultrasonication for 5 min. The resulting suspensions were then degased in a dessicator under agitation. The viscosities were measured in a rheometer (Gemini 200, Bohlin Instruments, England) in a serrated plate–plate configuration under controlled stress.

### *In situ* visualization of platelet response

The suspensions were maintained in closed pill glasses to prevent evaporation of the solvent. Rotating magnetic fields were applied using a 300-mT permanent neodymium magnet (Supermagnete, Switzerland) mounted on an electrical motor whose rotational speed is controlled by a power supply. The magnetic field strength was varied by changing the distance between the magnet and the sample and recorded at the sample location using a Gaussmeter (LakeShore, Cryotronics, USA). Color changes in the suspensions were recorded using a regular camera. Images were then extracted from the movies using the iMovie software (Apple, USA) and grey values were determined using the ImageJ program (NIH, USA). For each image, the grey values were normalized relative to the grey values of the edge of the pill glass, in order to erase any fluctuation in light intensity during recording. The data points were fitted using Matlab (MathWorks, USA).

### Casting

Porous gypsum substrates (Boesner, Switzerland) were prepared and dried in air before sample preparation. The casting of unidirectional or periodic structures was performed in the conditions and set-up described in refs 18 and 28.

### Characterization of the structures

Dried multi-scale structures were fractured and analyzed using a scanning electron microscope (SEM, Leo 1530, Zeiss, Germany). The degree of misalignment was semi-quantitatively described by the full width at half maximum (FWHM) of angular distributions extracted from the images using the ImageJ plugin MonogenicJ (http://bigwww.epfl.ch/demo/monogenic/) and a Matlab program[29]. X-ray diffraction (XRD) rocking curves were obtained with an Empyrean diffractometer (PANalytical, Germany) in reflection using Cu K$\alpha$ radiation ($\lambda = 1.5418\ \mathring{A}$) on porous structures infiltrated with a resin (SpeciFix-20, Struers, UK) and polished surfaces. The scan were performed around the (0012) crystallographic plane, corresponding to the basal plane of the platelets. The data was corrected using the software TexturePlus.

The distance between the platelets for suspensions of various volume fractions was estimated from epoxy–alumina samples with increasing equivalent amount of vertically aligned platelets.

## Results and discussion

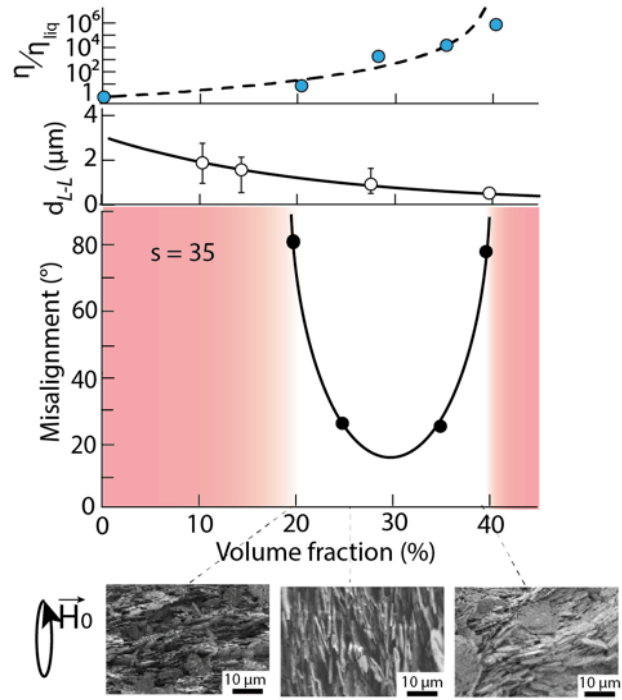

**Figure 2:** Variation in function of the volume fraction of platelets in %, of the relative viscosity $\frac{\eta}{\eta_{liq}}$, where $\eta_{liq}$ is the viscosity of the background, of the distance between the large faces of the platelets $d_{LL}$, as depicted on the cartoon figure 1d, and of the misalignment of the platelets (°). The dotted line in the viscosity plot corresponds to theoretical values obtained by applying the Krieger–Dougherty relation, whereas the filled black lines are guides to the eyes. The electron micrographs show cross-sections of MASC samples obtained under vertical rotating magnetic field of strength 50 mT and volume fractions of 20, 25 and 30 % respectively.

### Optimal concentration window

The alignment of the platelets, their assembly and alignment persistence during the MASC process and after final drying are highly dependent on the composition of the suspension and the inter-particle interactions within. An optimal composition of the suspension has to be determined for the platelets to easily orient them with the rotating magnetic fields but also to keep their assembly intact after drying (Fig. 2). The optimal composition of the suspension varies with the dimension and chemistry of the platelets and with the properties of the background fluid and the porous substrate, as these parameters influence the particles mobility. In this study, we study the case of alumina platelets of aspect ratio 30 that were coated with SPIONs and suspended in water with 5 wt% PVP.

An increase in platelet content inevitably leads to an increase in relative viscosity, and the evolution can be fitted with the Krieger–Dougherty relation to obtain the volume fraction where particles become jammed together, $v_m$:

$$\eta_r = \frac{\eta}{\eta_{liq}} = (1 - \frac{v}{v_m})^{-Bv_m} \tag{13}$$

where $\eta_r$ is the relative viscosity, defined as the ratio between the viscosity of the suspension and the viscosity of the background liquid, $v$ the volume fraction in particles, and *B* is the Einstein coefficient[25]. For the platelets considered here, a maximum $v_m$ of 40 vol% can be reached. Above this concentration, the assembly is jammed and exhibits solid-like behavior. This sets the highest volume fraction for the composition of the suspension.

When the volume fraction is decreased, the distances $d_{L-L}$ and $d_{l-l}$ between aligned platelets increases. This increases the volume available to the platelets for rotation. As a consequence, during the flow of the fluid from the suspension into the mold, the platelet can sink due to the combined action of capillary force and gravity. This sets the minimum volume fraction of particles within the slurry.

The minimum volume fractions $v_{min}^{opt}$ calculated using equation 15 give a value 16.8 vol% for these platelets. These volume fractions are comparable to the minimum concentration $v_{I-N}$ at which colloidal particles of the same aspect ratios start to form nematic phases due to packing entropy effects as described by the Onsager theory[25,26]. According to this theory, the coexistence of isotropic and nematic domains starts for $nL^3$ ~ 6.8, where *n* is the critical number density of platelets and *L* the longest dimension of the particle. The minimum volume fraction at which nematic domains form is then given by:

$$v_{I-N} = \frac{\pi n L^3}{4\frac{L}{l}} \tag{17}$$

Applying this to our particles, we found that $v_{I-N} = 17\%$ which is close to the experimental value, ranging between 15 and 20 vol%. For the latter case, the high polydispersity of the particles has to be taken into account, which is estimated to broaden the coexistence region of nematic and isotropic phases following a square scaling law [27,28].

To verify the expected window of alignment for our system, we looked at the microstructure of dried scaffolds prepared with increasing volume fractions in the initial slurry and a rotating magnetic field oriented for vertical alignment, which defines the optimum case for MASC. The FWHM values of the platelet assemblies were determine from electron micrographs. From those observations, an optimal range between 22 to 35 vol% can be determined for our experimental system (Fig. 2). The following results were all obtained using a slurry composed of 25 vol% for these platelets.

To summarize this part of our study, we demonstrated that in MASC the directed alignment is preserved only in a certain range of suspension volume fractions. This window can be estimated based on the distance between platelets and the strength of the capillary forces. Interestingly, the conditions for optimal suspension composition are not unique to the MASC process, but can be extended to any casting process with solvent removal.

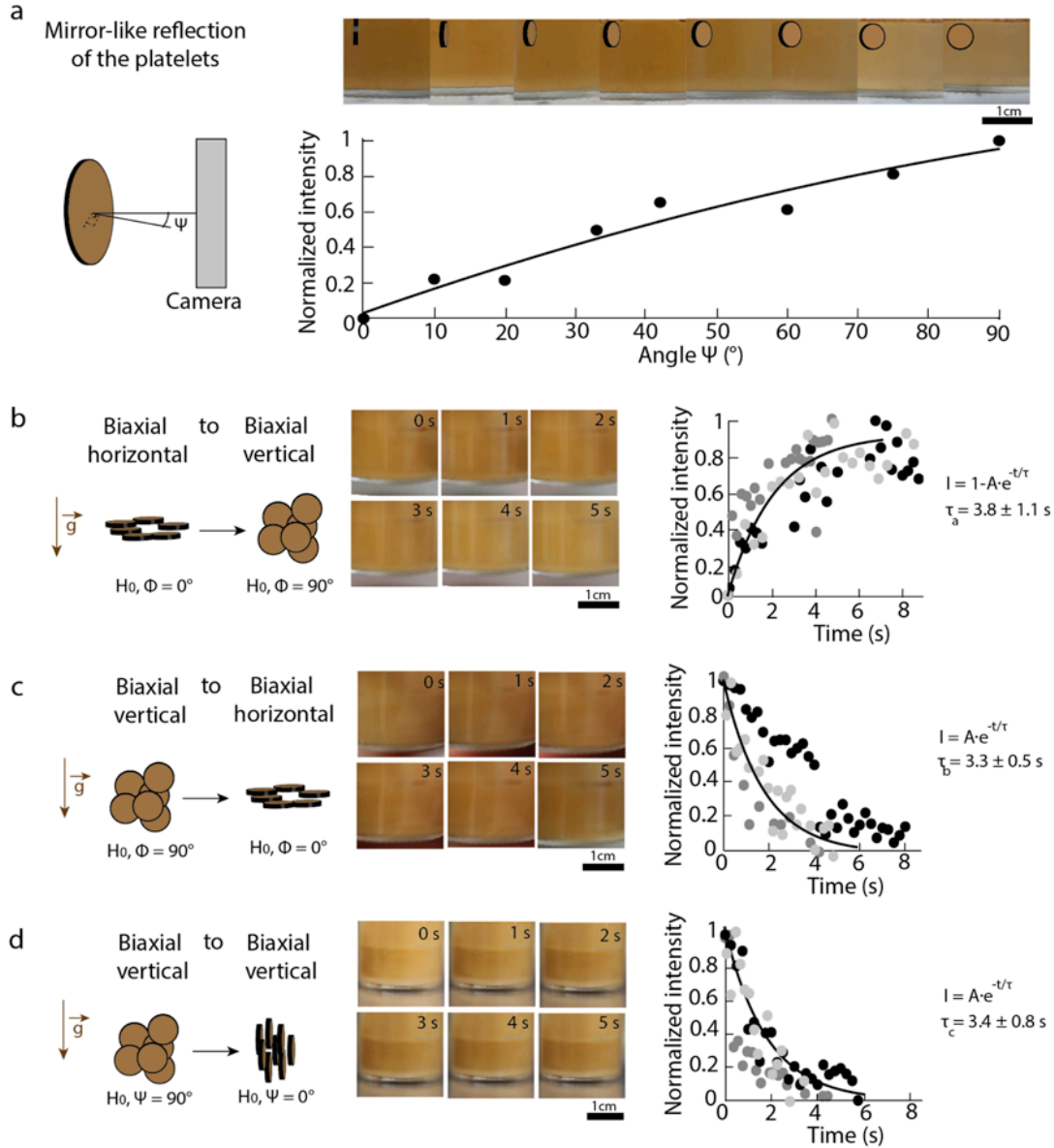

**Figure 3:** Determination of the platelets alignment time within the slurry. (a) Variation of the recorded normalized reflected intensity and corresponding optical micrographs as a function of the platelets' angle, $\psi$, as defined in the schematics. The black line corresponds to the fit obtained using the equation (14). (b), (c) and (d) depict the temporal response of the platelets submitted to various changes in rotating magnetic field orientations. The schematics on the left represent the position of the platelets in the suspension at time zero, just before changing the magnetic field orientation and after magnetic alignment. The optical images show the color change in the suspension composed of 25 vol% platelets in 5 wt% PVP aqueous solution and submitted to a magnetic field strength $H_0$ = 165 mT rotating at 1 Hz. The normalized intensity as a function of time is then fitted by an exponential decay law of the form $I = I_0 \cdot e^{-t/\tau}$. The time constant $\tau$ are averaged over the three set of data.

**Variation of the magnetic alignment with time**

During MASC, it is possible to change the platelets orientation as the jammed layer thickens by applying a stepwise temporal variation in rotating magnetic field direction. It is thus crucial to know what is the time necessary to align platelets and to compare it to the growth of the jammed layer. Our experiments show that this time does not depend significantly on the initial state and the alignment direction which is an important feature to build layered structures with layer-specific platelet orientation.

Considering the transient state during which the liquid is removed from the suspension, the minimum size of one layer with similarly oriented particles is directly related to the minimum time $\overline{\tau}$ required to change the orientation of the particles. $\overline{\tau}$ needs to be short enough compared to the time needed to grow a layer in order for the particles to have the same orientation within that layer. The layer thickness is a crucial parameter that influences the sample properties, such as energy absorption or crack deflection.

The motion of platelets with B in suspension can be directly visualized by optical microscopy, but only for extremely dilute system. However, in the case of high concentration, the color of the suspension changes with the orientation of the platelets, enabling thus the direct experimental assessment of these times $\overline{\tau}$. When the platelets in the suspension are facing the camera (angle $\psi$ = 90°), the reflected intensity is high and the colour is lighter than when the side of the platelets is in the direction of the camera (angle $\psi$ = 0°). This variation of the normalized intensity with the platelet's angle follows Fresnel's law: for specular reflections, the face of a platelet reflects the light in a mirror-like manner:

$$I(\psi) = a_0 + a_1 \cos(w\psi) + a_2 \sin(w\psi) \tag{14}$$

with $a_0$, $a_1$, $a_2$, and $w$, experimental constants.

The data points are fitted with the equation (14) with a correlation coefficient of $R^2$ = 0.95. This observation therefore allows us to analyse directly the recorded normalized intensity to follow the platelets' alignment and extract the minimum times $\overline{\tau}$.

Platelets can be rotated independently around two angles (named $\phi$ and $\psi$ in Fig. 1), but some of the torques are only present along one direction, two cases have to be considered. In the first case, when the magnetic fields are applied to rotate the particles around the angle $\phi$, the magnetic torque should counterbalance both the gravitational and the viscous torque (Fig. 3b). However, in the case of a variation around the angle $\psi$, only the magnetic and the viscous torques have to be taken into account, suggesting a faster temporal response (Fig. 3c,d). To evaluate $\overline{\tau}$, we measure the exponential decay time constants $\tau$, which corresponds to 63% of the final alignment. The minimum time $\overline{\tau}$ to reach 90% of the final alignment is then equal to $2.3\tau$. We consider thus three extreme cases (Fig. 3): from horizontal to vertical alignment (case b), from vertical to horizontal (case c), and between two vertical orientations (case d). The composition of the suspensions is kept constant, with 25 vol% of platelets of aspect ratio 30 suspended in 5 wt% PVP. The magnetic field applied has a strength of 160 mT and is rotating at a frequency of 1 Hz, which is above the critical frequency of individual platelets, determined at 0.65 Hz in average. Angle variations of $\psi = 90°$ are used as they induce the highest change in reflected light intensity. We can then assume that for smaller angles a smaller time will be needed. The time constant $\tau$ are extracted from the experimental data by fitting the values with an exponential decay law of the form $I = I_0 \cdot e^{-t/\tau}$. The times measured in the three compositions are $\tau_b$, $\tau_c$, and $\tau_d$ of 3.8 ± 1.1, 3.3 ± 0.5 and 3.4 ± 0.8 s respectively (Fig. 3b,c,d). These time constants are averaged using three independent suspensions and for experiments performed in the same conditions. The small decrease in time from configuration b to configuration c and d could be explained by the gravitational torque acting in favour of the horizontal alignment in c and not playing a role in d, respectively. But generally, the time needed to change the orientation of the platelets does not significantly vary with the configuration considered.

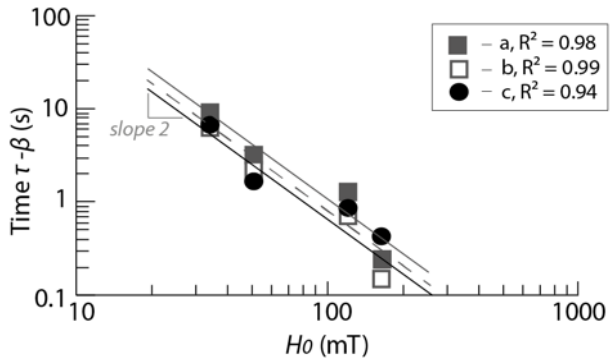

**Figure 4:** Variation of the constant of time $\tau - \beta$ for the four configurations b, c and d as defined in Fig.3, for increasing magnetic field strength.

The time necessary to align platelets can be further tuned by varying the magnetic response of the particles or the magnetic field strength applied to increase the magnetic torque. We show that as predicted in dilute suspensions, this time decreases with increasing magnetic field strength (Figure 4). Writing the torque balance of a platelets using the torques analytical expressions, the platelet angle is expressed by a non-trivial second order differential equation[17]. However, considering large variation of angles of 90° instead of infinitesimal variations, trends can be derived by writing the balance of torques on the platelet. In the cases b and c in Figure 3, the equilibrium is reached when

$T_m + T_v + T_g = 0$ (15)

while in the case d,

$T_m + T_v = 0$ (16)

Replacing the torques in (15) and (16) by their theoretical expressions [17], and considering their maximum, it appears that the time constants for the four configurations follow the same trend with the magnetic field strength $H_0$ $\tau = \frac{\alpha}{H_0^2} + \beta$, with $\alpha$ and $\beta$ experimental constants. As plotted in figure 4, the dependence of the alignment time on B in the concentrated slurries used for MASC is similar as obtained using equations describing the case of single platelets in dilute suspensions. Depending on the desired size of the sample and its distance with the magnet, the time $\bar{\tau}$ between each step should then be adjusted.

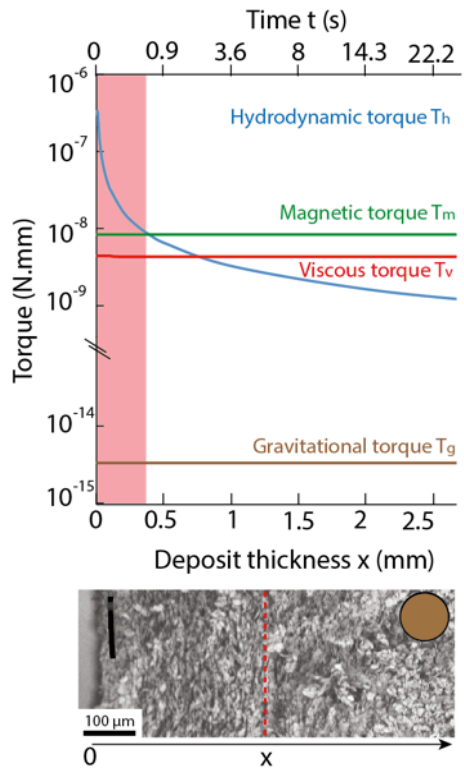

**Figure 5**: First instant of the casting: theoretical evolution of the torques applied to platelets during MASC as function of the casting time and deposit thickness. The high fluid flow in the first instant of the process leads to the formation of a horizontally aligned layer, visible in the SEM image of a cross-section of a vertically aligned dried sample prepared with 25 vol% platelets.

**Removal of the fluid**

Magnetic alignment enables the deliberate orientation of anisotropic platelets in suspension. However, during MASC, the removal of the fluid through the pores of the substrates introduces fluid flow that can disturb this alignment. In particular, at the first instants of the casting, this fluid flow is at its maximum leading to high hydrodynamic forces. These hydrodynamic forces guide the alignment of the platelets within the first layers. Using the equations given in (1), (2) and (3), we can calculate the maximum values of the various theoretical torques applied to the platelets during the casting. With a constant a = 0.534 mm.s$^{-0.5}$ and a cylindrical porous mould of diameter 10 mm, we calculate a fluid flow of $|Q(t)| = 20.959 \cdot t^{-0.5}$ in mm$^3$/s. The theoretical predictions show that the hydrodynamic torque is dominating the platelets orientation in the region close to the surface of the substrate (Figure 5). In this region, the hydraulic pressure gradient is very high due to the capillary forces from the porous mold, significantly increasing the drag force and hydrodynamic torque on the particles[29]. In consequence of this high flow, the platelets align horizontally at the surface of the gypsum, similarly to what is observed during the spreading of a liquid droplet onto a porous substrate[30,31]. This horizontal region spans up to 500 µm in the case of 200 nm-thick platelets, which shrinks down to 400 µm after taking into account the 20% linear shrinkage occurring during the final drying (value measured experimentally). This value is in good agreement with the thickness measured in electron micrographs for samples made in the same condition as used in the calculations above.

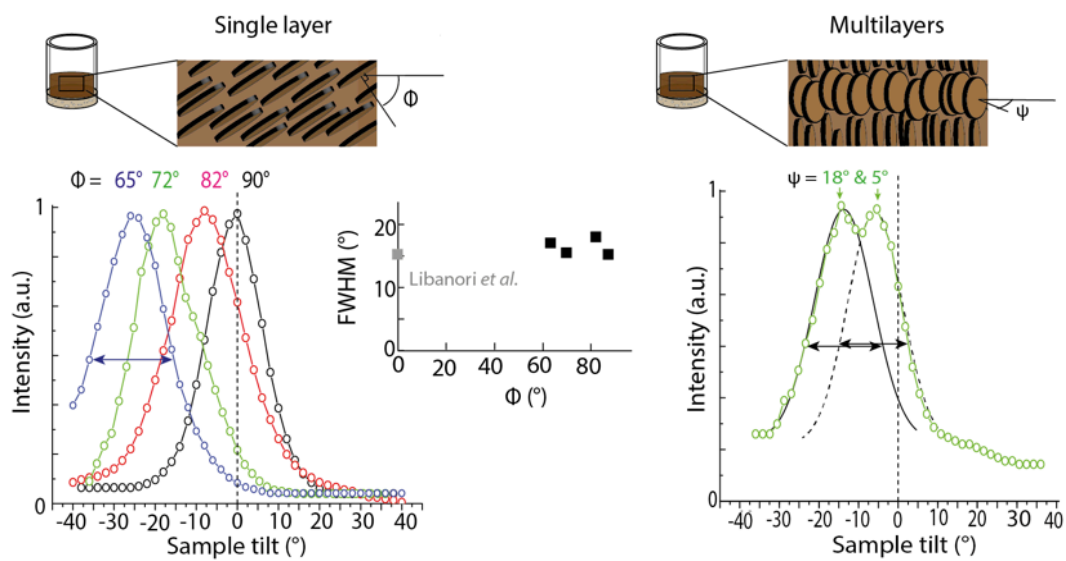

**Figure 6:** Control of the alignment using MASC in dried single layers with one specific platelet orientation (a) and in multi-layered structures with two pre-defined orientations (b) as measured using rocking curves. The insert in (a) represents the Full Width at Half-Max (FWHM) in function of the alignment angle Phi and is compared with the value obtained in reference 33.

**Control of the orientation**

Above the first layer of the casting, the alignment control obtain has an accuracy similar to that obtained with other processing techniques operating with lower volume fractions of platelets but for multiple alignment direction. This is a unique ability of MASC to fabricate accurately dense samples with controlled orientation of platelets.
The degree of alignment within each layer directly influences the property of each layer and in turns influence the properties of the final structure. We characterize the alignment degree by measuring the particles orientation distribution as a function of the angle $\phi$ between the magnetic field rotation plane and the sample surface using Rocking curves[32], a XRD-based texture analysis method.

The platelets orientation distribution of dried MASC-processed structures with unidirectional alignment show a narrow angle distribution with a FWHM of 17° on average over an angle $\phi$ from 0 to 25° (Fig. 6 a). This value is in good agreement with the degree of misalignment measured by image analysis (Fig 2) and is comparable to that obtained in epoxy-reinforced composites produced by magnetic alignment[33]. The higher misalignment degree in MASC in comparison with the one measured in epoxy composites[33] can either be due to the higher volume fraction (35%vol compare to 27%vol respectively) or to an additional misalignment introduced during the final drying step by the capillary forces.

In addition, structures with platelets orientation alternatively varying from $\phi$=5° and 18°, with a layer thickness of 200 µm, present two distinguishable peaks (Fig. 6b). The FWHM of each individual layers is similar to the one from a single layer, which suggest no influence of the variation of the alignment plane over time in the jammed layer.

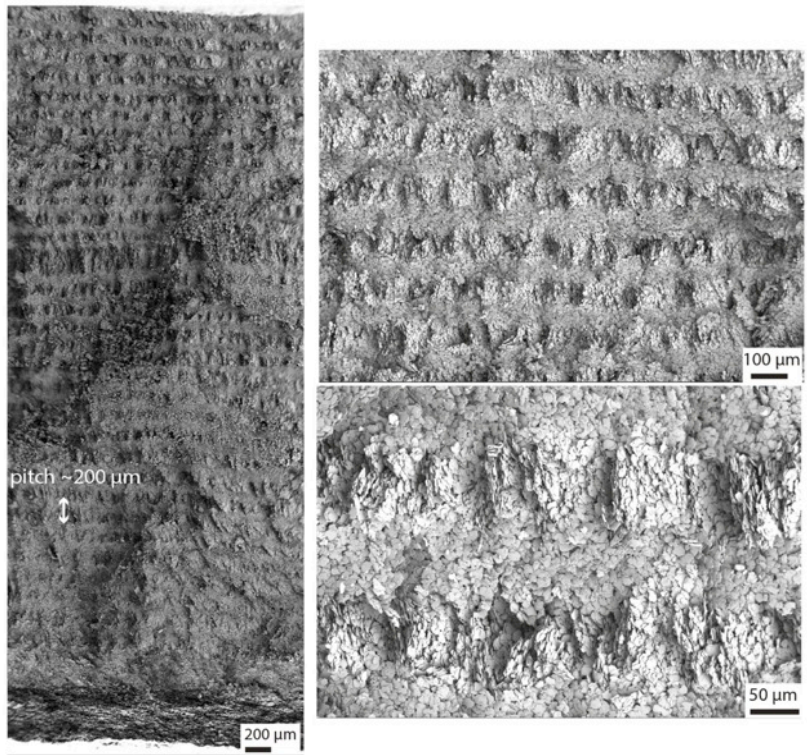

**Figure 7:** Structure obtained with decreasing the time between each layer such as approaching a periodic structure. One can note the horizontal layer at the bottom as predicted by the hydrodynamics. Electron micrographs of cross-sections of periodic microstructures build with time steps $\overline{\tau}_d = 30\ s$ and angle variation of $\psi = 90°$.

**Expansion of the design space**

By changing the dimensions of the platelets and controlling carefully their angle and the time required for successful alignment, a large range of microstructures with defined and tailored layer thickness and pitches can be manufactured. As an addition to previously published examples[14], we fabricated an almost periodic structure with vertically oriented platelets rotating along the angle Psi with an angle variation of 90° (Fig. 7). Knowing the casting parameters, the time between consecutive steps was varied in order to generate an almost constant layer thickness and pitch throughout the sample. This example further underlines the potential of the MASC method for building complex bio-inspired composites, enlarging the structural space to chemical diversity and property combinations.

**Conclusions**

The assembly of microstructures with arbitrarily defined orientations of anisotropic particles using MASC is affected by magnetic, hydrodynamic, viscous, steric and capillary forces developed during the assembly and drying stages of the process. Magnetic, hydrodynamic and viscous forces control the alignment dynamics of the anisotropic particles suspended in the liquid phase, whereas a balance between steric and capillary forces determine whether the obtained alignment can be preserved during the subsequent drying process. Understanding and quantifying the forces applied at different stages of the process enabled us to define guidelines and limitations of MASC in terms of suspension composition and the temporal evolution of the applied magnetic field. The particle volume fraction in the suspension should be sufficiently high to form an interlocking network that sterically resists the capillary forces that develop during complete removal of the liquid, but should be low enough to allow platelet movement with the rotating magnetic field. The time period under a fixed magnet orientation has to be higher than the minimum time required to align the platelets along that field direction. The lower boundary of the particle volume fraction window was found to be close to the concentration at which an isotropic–nematic phase transition occurs, driven by inter-particle steric interactions. The upper boundary is defined by the significant increase in platelet interactions at high volume fractions. The alignment time of the particles depends on the applied magnetic field and can be quantified using simple scaling laws obtained for diluted systems.

This first comprehensive description of the MASC process should enable its application for producing a large variability of structures. Similar formulation guidelines could be determined for any other type of anisotropic building blocks, taking into account the relevant chemistries, shapes and dimensions. Importantly, except for the hydrodynamic force generated by the fluid removal, these guidelines can be used for any process involving solvent removal. The ability demonstrated here to finely tailor the structural internal features of centimeter-sized samples should open up new possibilities for designing and producing unusual composites exhibiting new and various functionalities.

**Acknowledgments**

We acknowledge Nik Kranzlin for his help with the XRD experiments and funding from ETH Zürich and from the Swiss National Science Foundation (grant 200020_146509). H.L.F and F.B. designed the research, H.L.F performed the experiments and H.L.F. and F.B. prepared the manuscript. H.L.F, F.B. and A.R.S discussed the results and their implications and revised the manuscript.

**Supplementary Figure**

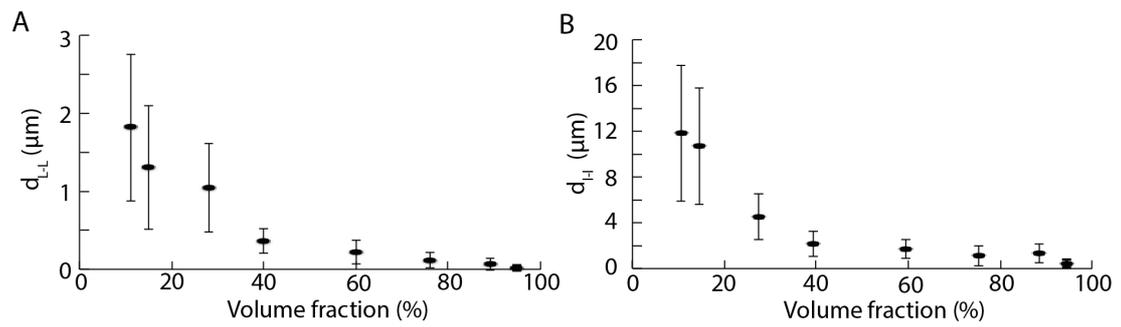

**Supplementary Figure: A,B)** Experimental values of the interplatelet distances for the particles of aspect ratio 30.